\documentclass{article}
\usepackage{graphicx}
\usepackage{amsmath}

\input{tcilatex}

\begin{document}

\title{\textbf{2T Physics, Scale Invariance}\\
\textbf{and Topological Vector Fields}}
\author{W. Chagas-Filho \\
Physics Department, Federal University of Sergipe, Brazil}
\maketitle

\begin{abstract}
We construct, in classical two-time physics, the necessary structure for the
most general configuration space formulation of quantum mechanics containing
gravity in d+2 dimensions. This structure is composed of a symmetric
Riemannian metric tensor and of a vector field that defines a section of a
flat U(1) bundle over space-time. This construction is possible because of
the existence of a finite local scale invariance of the Hamiltonian and
because two-time physics contains, at the classical level, a local
generalization of the discrete duality symmetry between position and
momentum that underlies the structure of quantum mechanics.
\end{abstract}

\section{Introduction}

The symmetry transformations of a classical action functional describing a
physical system can be divided into four types. The most common type
consists of the rigid (global) infinitesimal symmetry transformations. These
are infinitesimal transformations of the dynamic variables, and possibly of
the auxiliary variables that appear in the action functional, parametrized
by constant arbitrary infinitesimal parameters.

Also quite common are the local infinitesimal symmetry transformations,
which compose the second type. These are infinitesimal transformations of
the dynamic variables, and auxiliary variables, parametrized by arbitrary
infinitesimal parameters which depend on the manifold point where the
transformations are performed.

The last two types of symmetry transformations are not infinitesimal
transformations and correspond to the finite rigid and finite local symmetry
transformations. These are in some cases rather subtle symmetry
transformations because their finite character is related with topological
aspects of the underlying manifold. And this relation is important because
the topological aspects of a manifold are related to the non trivial
diffeomorphic-covariant representations of the Heisenberg algebra over the
manifold [1]. We can then use the finite symmetry transformations on a
certain topologically non-trivial manifold to investigate the general
structure of quantum mechanics.

In this paper we are interested in the Lorentz $SO(d,2)$ invariance. This
invariance manifests itself as conformal invariance of the scalar
relativistic massless particle in a $d$ dimensional Minkowski space only if
a compactification of the space-time is assumed [2,3]. $SO(d,2)$ turns out
to be the isometry of $(d+1)$-dimensional Anti de Sitter (AdS) space if a
slightly different compactification of space-time is assumed [2]. These two
slightly different compactifications then reveal that the $d$-dimensional
Minkowski space is the border of the $(d+1)$ dimensional AdS space, an
observation that is the cornerstone of the AdS/CFT conjecture [4]. It is
then important to understand other possible ramifications of the Lorentz $%
SO(d,2)$ invariance.

$SO(d,2)$ is also the rigid symmetry of two-time (2T) physics [5-11], where
it appears as a consequence of the first class Hamiltonian constraints. From
the point of view of 2T physics the known fundamental gravitational and
gauge interactions in $d$ dimensions are all embedded in a $d+2$ dimensional
flat Minkowski space with two timelike dimensions. From this point of view
the fundamental interactions display higher dimensional space-time
symmetries that otherwise would remain hidden. In the current formulation of
two-time physics compactification of the $(d+2)$-dimensional Minkowski space
is avoided by considering only infinitesimal rigid $SO(d,2)$
transformations. In particular, only infinitesimal rigid scale and special
conformal transformations are defined. The local versions of these two
infinitesimal transformations, together with diffeomorphism invariance,
compose the local $Sp(2,R)\sim SO(1,2)$ gauge invariance of two-time physics.

However, despite avoiding the consideration of the implications of
space-time compactification, topological considerations necessarily arise in
consequence of the nontrivial configuration space topology induced by the
first class constraints of 2T physics. These topological considerations have
a fundamental origin. Quantum dynamics requires the definition of a
Riemannian metric structure on configuration space, whose determinant
directly specifies the normalization of position eigenstates in order to
ensure the correct covariant properties of the Heisenberg algebra
representations under diffeomorphisms of the configuration manifold [1]. In
addition, due to the local arbitrariness in the phase of position
eigenstates, 
\begin{equation}
\mid \tilde{x}\rangle =\exp \{\frac{i}{\hbar }\beta (x)\}\mid x\rangle 
\tag{1.1}
\end{equation}
a flat U(1) bundle is always associated to any such representation of the
Heisenberg algebra [1]. In the case of a simply connected manifold, this
flat U(1) bundle may always be globally trivialized over the entire
configuration manifold $M$, thereby corresponding to the ordinary trivial
representation of the Heisenberg algebra. However, for configuration spaces
of non trivial mapping class group $\pi _{1}(M)$, an infinity of
inequivalent representations becomes possible, being labelled by the non
trivial holonomies of the flat U(1) bundle around the noncontractile cycles
in the configuration manifold [1]. This last situation is exactly the one we
have in 2T physics. This is because the Hamiltonian constraints require, for
consistency, that the origin of phase space be removed. This induces a
non-trivial configuration space topology which will then require the
presence, in the quantized 2T theory, of a vector field of vanishing
strength tensor associated to the flat U(1) bundle which will characterize
the inequivalent representations of the Heisenberg algebra. The search for a
naturally induced metric structure in the $d+2$ dimensional space of 2T
physics, and the construction of a classical 2T action with a background
vector field of vanishing strength tensor, are the subjects of this paper.

A natural geometrical interpretation [20] of gauge fields is to identify the
vector potentials $A_{M}$ with the connection coefficients of the principal
fiber space whose base is Riemannian space-time, the fiber being a finite
gauge Lie group $G$. In this case, the stress tensor $F_{MN}$ of the gauge
field becomes the curvature tensor of the fiber space. For a flat U(1)
bundle, $F_{MN}=0.$ \ An approach to the introduction of background
gravitational and gauge fields in 2T physics was first presented in [11]. In
[11], the linear realization of the $Sp(2,R)$ gauge algebra of two-time
physics is required to be preserved when background gravitational and gauge
fields come into play. To satisfy this requirement, the gravitational field
must satisfy a homothety condition [11], while in the absence of
gravitational fields the gauge field $A_{M}(X)$ must satisfy the conditions
[11] 
\begin{equation}
X.A(X)=0  \tag{1.2a}
\end{equation}
\begin{equation}
\partial _{M}A^{M}(X)=0  \tag{1.2b}
\end{equation}
\begin{equation}
(X.\partial +1)A_{M}(X)=0  \tag{1.2c}
\end{equation}
which were first proposed by Dirac [12] in 1936. Dirac proposed these
conditions as subsidiary conditions to describe the usual 4-dimensional
Maxwell electrodynamic theory as a theory in 6 dimensions which
automatically displays $SO(4,2)$ symmetry.

Dirac's conditions (1.2) are a reflex of a hidden fundamental Sp(2,R)
symmetry in Maxwell's electrodynamics. This can be seen as follows. If we
recall that in the topologically trivial [1] transition to quantum mechanics
we can substitute $X^{M}\rightarrow X^{M}$ and $P_{M}\rightarrow i\hbar 
\frac{\partial }{\partial X^{M}}$, we can construct a semi-classical
approximation where derivatives with respect to $X^{M}$ are substituted by $%
P_{M}$ and rewrite Dirac's conditions (1.2) in the form 
\begin{equation}
X.A(X)=0  \tag{1.3a}
\end{equation}
\begin{equation}
P.A(X)=0  \tag{1.3b}
\end{equation}
\begin{equation}
(X.P+1)A_{M}(X)=0  \tag{1.3c}
\end{equation}
For an electrodynamic vector field, the condition for the closure of the $%
Sp(2,R)$ gauge algebra of 2T physics is [11] 
\begin{equation}
X^{M}F_{MN}=X^{M}(\frac{\partial A_{N}}{\partial X^{M}}-\frac{\partial A_{M}%
}{\partial X^{N}})=0  \tag{1.4}
\end{equation}
Using again our semi-classical approximation, condition (1.4) becomes 
\begin{equation}
(X.P+1)A_{N}=P_{N}(X.A)  \tag{1.5}
\end{equation}
We see from (1.5) that the $Sp(2,R)$\ closure condition (1.4) leads to
Dirac's condition (1.3c) only if we first impose condition (1.3a). When this
is done, the rigid $SO(4,2)$ invariance of electrodynamics is the reflex of
a local $Sp(2,R)$ invariance. But for the case we are interested here,
namely that of a topological vector field associated to a flat U(1) bundle, $%
F_{MN}=0,$ and so condition (1.4) is trivially satisfied. Therefore, in the
case of a topological vector field, even if we impose condition (1.3a)
first, Dirac's condition (1.3c) can not be reached. On the contrary, if we
impose (1.3c) first, then (1.3a) can not be reached. We must then search for
an alternative set of conditions on the vector field if we want our
topological 2T action to display local $Sp(2,R)$ and consequently rigid $%
SO(d,2)$ invariances.

It is important to find this correct set of subsidiary conditions on the
topological vector field. As demonstrated in [1], vector fields of vanishing
strength tensor play a fundamental role in the generalization of quantum
mechanics in the position representation. In this generalization, this kind
of vector field also necessarily appears, together with the determinant of
the Riemannian metric tensor, in the most general expression of the position
matrix elements for self-adjoint momentum operators in configuration spaces
with non-trivial topology, 
\begin{equation*}
\langle X\mid \hat{P}_{M}\mid X^{\prime }\rangle =\frac{i\hbar }{G^{1/4}(X)}%
\frac{\partial }{\partial X^{M}}[\frac{1}{G^{1/4}(X)}\delta ^{d}(X-X^{\prime
})]
\end{equation*}
\begin{equation}
+\frac{1}{\sqrt{G(X)}}A_{M}(X)\delta ^{d}(X-X^{\prime })  \tag{1.6}
\end{equation}
where $G(X)=\det G_{MN}(X)$. We conclude from this generalization of quantum
mechanics that vector fields of vanishing strength tensor in topologically
non-trivial spaces will play an important role in the also important process
of further unifying general relativity and quantum mechanics beyond the
configuration space treatment exposed in [1].

However, as we saw above, the conditions (1.2) obtained in [11], or their
semi-classical approximations (1.3), are not guaranteed to be valid for
vector fields of vanishing strength tensor. Therefore these conditions can
not be used in the process of accommodating gravity into quantum mechanics
in the higher dimensional space-time of 2T physics. The set of conditions on
the vector field we obtain in this paper has a different nature from that of
the set (1.2). While the set (1.2) is formed from subsidiary kinematical
conditions with no special significance, the set of conditions we obtain
here has a more fundamental origin because it is formed by the first class
Hamiltonian constraints for 2T physics with a topological vector field. This
will be explicitly verified in section four, where we show that these first
class constraints compose the correct conserved Hamiltonian Noether charge
associated to local $Sp(2,R)$ invariance in the presence of the topological
vector field.

The formulation of 2T physics with vector fields we present in this paper
has implications, some of which are now being investigated, in the
non-relativistic and relativistic quantum mechanics, in $d-1$ dimensions and 
$d$ dimensions respectively, of physical systems enjoying local
infinitesimal conformal $SO(1,2)\sim Sp(2,R)$ symmetry and/or global
infinitesimal Lorentz $SO(d,2)$ symmetry$.$ The list of these systems starts
with the free massive non-relativistic particle and ends with black holes,
passing through the harmonic oscillator, the Hydrogen atom, the de Sitter
and Anti de Sitter spaces, and contains all the dynamic systems that have a
unified description given by 2T physics. Our formulation of 2T physics with
vector fields can also lead to interesting insights into the implications of
the wave-particle duality on the general structure of quantum mechanics and
provide a formulation of quantum mechanics with a single time where position
and momentum are explicitly treated as locally indistinguishable variables
[21]. Recall that the results in [1] are valid for configuration space only.
Here the first class constraint structure of two-time physics, which is what
ultimately requires a metric with two time-like dimensions, also requires
the origin of phase space\textbf{\ }to be removed [3]. This creates a
non-trivial phase space topology, with inequivalent diffeomorphic covariant
representations of the Heisenberg algebra over the configuration space
(viewed as part of the phase space), which are all classified in terms of
vector fields with a vanishing second-rank antisymmetric strength tensor. In
this paper we study this situation at a semi-classical level, present a
Hamiltonian formulation of 2T physics with such a kind of vector fields, and
show that the action we compute has a rigid infinitesimal $SO(d,2)$
invariance. We also show that our action has a local infinitesimal
invariance which generalizes the local $Sp(2,R)$ invariance in the presence
of the vector field and compute the corresponding conserved Hamiltonian
Noether charge. These results can be used as the basis for a formulation of
quantum mechanics which naturally accommodates gravity in higher dimensions
based on the construction described in [1]. They also suggest that a
momentum space version of the results in [1] is straightforward.

It has been known for some time [20] that the gravitational field, regarded
as a gauge field, can correspond to several symmetry groups: 1) the general
covariant group; 2) the local Lorentz group; and 3) the group of scale
transformations of the interval. In the first case the properties of the
gravitational field are determined by the properties of the metric tensor,
and this gives the usual Einstein theory. In the second case, they are
determined by the properties of the Ricci connection coefficients, and this
leads to field equations of fourth order. In the third case it is assumed
that the source of the field is the trace of the energy-momentum tensor and
that the carriers are scalar particles [20]. Consequently, the approach
based on gauge symmetries can lead to more general theories than that of
Einstein. In this paper we are somewhat in the context of the third point of
view. This is because Lorentz $SO(d,2)$ invariance manifests itself \ as
conformal invariance of the relativistic scalar massless particle action in $%
d$ dimensions, and the 2T physics action is the higher dimensional
generalization of the scalar massless particle action to $d+2$ dimensions.
For these reasons, in the next section we examine the rigid and local
symmetries of the massless particle action. We present a finite local scale
invariance of the particle's Hamiltonian that induces a transformation of
the position coordinates which in this paper we are inclined to interpret as
the classical correspondent of the quantum local phase transformation (1.1)
of the position eigenstates. We also show how we can use this local scale
invariance of the massless particle Hamiltonian to derive the classical
analogues of the Snyder commutators [18], which were derived in 1947 in a
projective geometry approach to the de Sitter space in the momentum
representation.

In section three we review the construction of the 2T physics action and
explicitly display its rigid and local infinitesimal symmetries. We compute
the conserved Hamiltonian Noether charge and show that the finite local
scale invariance we found for the massless particle has a simple and natural
extension in 2T physics. Then we show how we can use this finite local scale
invariance of the 2T Hamiltonian to induce a Riemannian metric structure in $%
d+2$ dimensions.

In section four we construct an action functional for 2T physics in the
background of a vector field of vanishing strength tensor. We display its
rigid infinitesimal Lorentz $SO(d,2)$ invariance and compute the conserved
Hamiltonian Noether charge in the presence of the vector field. We find that
this conserved charge is composed of the original first class constraints of
2T physics complemented with three first class constraints which involve the
vector field and the canonical variables. These last three first class
constraints must be used in the place of conditions (1.3) when $F_{MN}=0$.
We also show how the Riemannian metric structure we found in the absence of
the vector field in section three is preserved in the presence of the vector
field. We conclude that we have found, already in classical 2T physics, the
fundamental necessary ingredients for the topologically non trivial
construction of quantum mechanics described in [1], that is, a Riemannian
metric structure and a vector field of vanishing strength tensor. We were
able to do this because 2T physics has a classical local $Sp(2,R)$ invariant
generalization of the discrete duality symmetry between position and
momentum that underlies the structure of quantum mechanics. Other concluding
remarks appear in section five.

\section{Massless Relativistic Particles}

Before considering topological aspects in 2T physics, it is instructive to
consider these aspects in massless scalar particle theory. A massless scalar
relativistic particle in a $d$-dimensional Minkowski space with signature $%
(d-1,1)$ is described by the Lagrangian action 
\begin{equation}
S=\frac{1}{2}\int d\tau \lambda ^{-1}\dot{x}^{2}  \tag{2.1}
\end{equation}
where $x^{\mu }(\tau )$ are the position coordinates, $\lambda (\tau )$ is
an auxiliary variable and a dot denotes derivatives with respect to the
parameter $\tau $. Action (2.1) is invariant under the local infinitesimal
reparametrizations 
\begin{equation}
\delta x_{\mu }=\epsilon (\tau )\dot{x}_{\mu }\text{ \ \ \ \ }\delta \lambda
=\frac{d}{d\tau }[\epsilon (\tau )\lambda ]  \tag{2.2}
\end{equation}
and therefore describes gravity on the world-line. Action (2.1) is also
invariant under the following rigid infinitesimal transformations.
Poincar\'{e} transformations 
\begin{equation}
\delta x^{\mu }=a^{\mu }+\omega _{\nu }^{\mu }x^{\nu }\text{ \ \ \ \ }\delta
\lambda =0  \tag{2.3}
\end{equation}
where $\omega _{\mu \nu }=-\omega _{\nu \mu }$ is a constant matrix, under
the scale transformations 
\begin{equation}
\delta x^{\mu }=\alpha x^{\mu }\text{ \ \ \ \ }\delta \lambda =2\alpha
\lambda  \tag{2.4}
\end{equation}
where $\alpha $ is a constant, and under the conformal transformations 
\begin{equation}
\delta x^{\mu }=(2x^{\mu }x^{\nu }-\eta ^{\mu \nu }x^{2})b_{\nu }\text{ \ \
\ \ }\delta \lambda =4\lambda x.b  \tag{2.5}
\end{equation}
where $b_{\mu }$ is a constant vector. Finite conformal transformations,
given by [3] 
\begin{equation}
\tilde{x}^{\mu }=\frac{x^{\mu }+b^{\mu }x^{2}}{1-2b.x+b^{2}x^{2}}  \tag{2.6a}
\end{equation}
\begin{equation}
\tilde{\lambda}=\frac{\lambda }{(1-2b.x+b^{2}x^{2})^{2}}  \tag{2.6b}
\end{equation}
are not globally defined, and to be well defined require a compactification
of the $d$-dimensional Minkowski space by including the points at infinity.
A possible compactification is the ``quadric''\ described in [2]. In this
paper we will not assume such a compactification, and therefore finite
conformal transformations of the type (2.6) will not be considered as
symmetries of action (2.1).

Although action (2.1) is not invariant under the finite conformal
transformations (2.6), it is invariant under the finite scale transformation
[3] 
\begin{equation*}
\tilde{x}^{\mu }=\exp \{\beta \}x^{\mu }\text{ \ \ \ \ \ }\tilde{\lambda}%
=\exp \{2\beta \}\lambda
\end{equation*}
where $\beta $ is a constant parameter. But action (2.1) is not invariant
under the local infinitesimal scale transformation $\delta x_{\mu }=\beta
(\tau )x_{\mu }$ , $\delta \lambda =2\beta (\tau )\lambda $ , nor under the
finite local scale transformation $\tilde{x}_{\mu }=\exp \{\beta (\tau
)\}x_{\mu }$ , $\tilde{\lambda}=\exp \{2\beta (\tau )\}\lambda $. As we will
see below, although finite local scale transformations are not symmetries of
action (2.1), they are symmetries of the corresponding canonical
Hamiltonian. This will turn out to be related to the appearance, in massless
particle theory, of the classical analogues of the old Snyder commutators,
derived for the de Sitter space in the momentum representation.

As a consequence of the infinitesimal invariances (2.3), (2.4) and (2.5) of
action (2.1) we can define in space-time the following vector field 
\begin{equation}
V=a^{\mu }P_{\mu }-\frac{1}{2}\omega ^{\mu \nu }M_{\mu \nu }+\alpha D+b^{\mu
}K_{\mu }  \tag{2.7}
\end{equation}
with generators 
\begin{equation}
P_{\mu }=p_{\mu }  \tag{2.8a}
\end{equation}
\begin{equation}
M_{\mu \nu }=x_{\mu }p_{\nu }-x_{\nu }p_{\mu }  \tag{2.8b}
\end{equation}
\begin{equation}
D=x.p  \tag{2.8c}
\end{equation}
\begin{equation}
K_{\mu }=2x_{\mu }x.p-x^{2}p_{\mu }  \tag{2.8d}
\end{equation}
$P_{\mu }$ generates translations in space-time,$\ M_{\mu \nu }$ is the
generator of Lorentz transformations, $D$ is the generator of space-time
dilatations and $K_{\mu }$ generates conformal transformations. These
generators define the algebra 
\begin{equation*}
\{M_{\mu \nu },M_{\lambda \rho }\}=\eta _{\nu \lambda }M_{\mu \rho }+\eta
_{\mu \rho }M_{\nu \lambda }-\eta _{\nu \rho }M_{\mu \lambda }-\eta _{\mu
\lambda }M_{\nu \rho }
\end{equation*}
\begin{equation*}
\{M_{\mu \nu },P_{\lambda }\}=\eta _{\mu \lambda }P_{\nu }-\eta _{\nu
\lambda }P_{\mu }\text{ \ \ \ }\{M_{\mu \nu },K_{\lambda }\}=\eta _{\nu
\lambda }K_{\mu }-\eta _{\lambda \mu }K_{\nu }
\end{equation*}
\begin{equation*}
\{D,P_{\mu }\}=P_{\mu }\text{ \ \ \ }\{D,K_{\mu }\}=-K_{\mu }\text{ \ \ \ }%
\{D,D\}=0
\end{equation*}
\begin{equation*}
\{K_{\mu },P_{\nu }\}=2(\eta _{\mu \nu }D+M_{\mu \nu })
\end{equation*}
\begin{equation}
\{D,M_{\mu \nu }\}=\{P_{\mu },P_{\nu }\}=\{K_{\mu },K_{\nu }\}=0  \tag{2.9}
\end{equation}
computed in terms of the Poisson brackets 
\begin{equation}
\{p_{\mu },p_{\nu }\}=\{x_{\mu },x_{\nu }\}=0\text{ \ \ \ }\{x_{\mu },p_{\nu
}\}=\eta _{\mu \nu }  \tag{2.10}
\end{equation}
The algebra (2.9) is the conformal space-time algebra. The scalar massless
particle theory defined by action (2.1) is a conformal theory in $d$
dimensions.

Conformal invariance in $d$ dimensions is isomorphic to Lorentz invariance
in $d+2$ dimensions. By defining [3] 
\begin{equation}
L_{\mu \nu }=M_{\mu \nu }  \tag{2.11a}
\end{equation}
\begin{equation}
L_{\mu d}=\frac{1}{2}(P_{\mu }+K_{\mu })  \tag{2.11b}
\end{equation}
\begin{equation}
L_{\mu (d+1)}=\frac{1}{2}(P_{\mu }-K_{\mu })  \tag{2.11c}
\end{equation}
\begin{equation}
L_{d(d+1)}=D  \tag{2.11d}
\end{equation}
the conformal algebra (2.9) can be put in the standard form 
\begin{equation}
\{L_{MN},L_{RS}\}=\eta _{MR}L_{NS}+\eta _{NS}L_{MR}-\eta _{MS}L_{NR}-\eta
_{NR}L_{MS}  \tag{2.12}
\end{equation}
with $M,N=0,1,...,d,d+1$ and $\eta _{MN}=diag(-1,+1,...,+1,-1)$. This shows
that there are hidden dimensions in scalar massless particle theory. In the
next section we will use these hidden dimensions to generalize the
world-line action (2.1) to a more symmetric theory in a $(d+2)$-dimensional
space-time.

Lagrangian mechanics is contained in Hamiltonian mechanics [13]. To be more
general we must pass to the Hamiltonian formalism. In the transition to this
formalism action (2.1) gives the canonical momenta 
\begin{equation}
p_{\lambda }=0  \tag{2.13}
\end{equation}
\begin{equation}
p_{\mu }=\frac{\dot{x}_{\mu }}{\lambda }  \tag{2.14}
\end{equation}
and the canonical Hamiltonian 
\begin{equation}
H=\frac{1}{2}\lambda p^{2}  \tag{2.15}
\end{equation}
Equation (2.13) is a primary constraint [14]. Introducing the Lagrange
multiplier $\xi (\tau )$ for this constraint we can write the Dirac
Hamiltonian 
\begin{equation}
H_{D}=\frac{1}{2}\lambda p^{2}+\xi p_{\lambda }  \tag{2.16}
\end{equation}
Requiring the dynamic stability of constraint (2.13), $\dot{p}_{\lambda
}=\{p_{\lambda },H_{D}\}=0$, we obtain the secondary constraint 
\begin{equation}
\phi =\frac{1}{2}p^{2}\approx 0  \tag{2.17}
\end{equation}
Constraints (2.13) and (2.17) have vanishing Poisson bracket, being
therefore first-class constraints [14]. The gauge transformations generated
by $\phi $ are discussed below. Constraint (2.13) generates translations in
the arbitrary variable $\lambda (\tau )$ and can be dropped from the
formalism.

In equation (2.17) we introduced [15] the \textbf{weak equality symbol }$%
\approx $. This is to emphasize that constraint $\phi $ is numerically
restricted to be zero in the subspace of phase space where the canonical
momentum satisfies equation (2.17), but $\phi $ does not identically vanish
throughout phase space. In particular, it has nonzero Poisson brackets with
the canonical positions. More generally, two functions $F$ and $G$ that
coincide on the submanifold of phase space defined by the constraints are
said to be \textbf{weakly equal }over phase space and one writes $F\approx G$%
. On the other hand, an equation that holds throughout phase space, and not
just on the submanifold defined by the constraint equations, is called 
\textbf{strong}, and the usual equality symbol is used in that case. It can
be demonstrated that, in general [15] 
\begin{equation}
F\approx G\Leftrightarrow F-G=c_{i}(x,p)\phi _{i}  \tag{2.18}
\end{equation}
where $\phi _{i}$ denote the constraints.

Equation (2.17) can be treated as a constraint only if the points with $%
p_{0}=p_{1}=...=p_{d-1}=0,$ corresponding to the trivial representation of
the Poincar\'{e} group, are excluded from phase space [3]. From the
definition of the canonical momentum (2.14) the points with $%
x_{0}=x_{1}=...=x_{d-1}=0$ must also be excluded for consistency. This
introduces a non-trivial phase space topology and makes a scalar massless
relativistic particle similar to the non-relativistic charge-monopole system
[3,16]. Due to this non trivial phase space topology, a flat U(1) bundle
will necessarily be present in the quantized massless particle theory.

To further develop the Hamiltonian formalism, we write action (2.1) in the
form 
\begin{equation}
S=\int_{\tau _{i}}^{\tau _{f}}d\tau (\dot{x}.p-\frac{1}{2}\lambda p^{2}) 
\tag{2.19}
\end{equation}
If we solve the equation of motion for $p_{\mu }$ that follows from (2.19)
and insert the result back in it, we recover action (2.1). Constraint (2.17)
generates the local infinitesimal transformation 
\begin{equation}
\delta x_{\mu }=\epsilon (\tau )\{x_{\mu },\phi \}=\epsilon (\tau )p_{\mu } 
\tag{2.20a}
\end{equation}
\begin{equation}
\delta p_{\mu }=\epsilon (\tau )\{p_{\mu },\phi \}=0  \tag{2.20b}
\end{equation}
\begin{equation}
\delta \lambda =\dot{\epsilon}(\tau )  \tag{2.20c}
\end{equation}
under which action (2.19) transforms as 
\begin{equation}
\delta S=\int_{\tau _{i}}^{\tau _{f}}d\tau \frac{d}{d\tau }(\epsilon \phi ) 
\tag{2.21}
\end{equation}
Since the interval $(\tau _{i},\tau _{f})$ is arbitrary, we see that action
(2.19) is invariant under transformations (2.20), and that the quantity $%
Q=\epsilon \phi $ can be interpreted as the conserved Hamiltonian Noether
charge or as the generator of the local transformations (2.20), depending on
wether the equations of motion are satisfied or not [17]. This particular
aspect of the quantity $Q$ will be used as a consistency check when we
introduce vector fields in 2T physics below.

The most general physically permissible motion should allow for an arbitrary
gauge transformation to be performed while the system is dynamically
evolving in time [15]. Since the dynamic time evolution of a physical system
is governed by its Hamiltonian, this arbitrary gauge transformation must
leave the Hamiltonian invariant. In the case of the scalar relativistic
massless particle, parametrized by $\tau $, we point out that the
Hamiltonian (2.15) is invariant under the finite local scale transformations 
\begin{equation}
\tilde{p}_{\mu }=\exp \{-\beta (\tau )\}p_{\mu }  \tag{2.22a}
\end{equation}
\begin{equation}
\tilde{\lambda}=\exp \{2\beta (\tau )\}\lambda  \tag{2.22b}
\end{equation}
where $\beta (\tau )$ is an arbitrary scalar function. From equation (2.14)
for the canonical momentum we find that $x^{\mu }$ transforms as 
\begin{equation}
\tilde{x}^{\mu }=\exp \{\beta (\tau )\}x^{\mu }  \tag{2.22c}
\end{equation}
when $p_{\mu }$ transforms as in (2.22a). The finite local scale
transformation (2.22) is a symmetry in phase space but, as we saw above, it
breaks down if we try a transition to configuration space. It is
interesting, in the case when $\beta (\tau )=\beta (x(\tau ))$, to try to
relate, using the correspondence principle, the local scale transformation
(2.22c) of the position variables with the local phase transformation (1.1)
of the position eigenstates. Gravity and the flat U(1) bundle would then be
related by finite local scale invariance. We will not consider this point
here.

Consider now the bracket structure that transformations (2.22a) and (2.22c)
induce in phase space. The following calculations are an improved, more
rigorous version, of the ones which appear in [19]. Retaining only the
linear terms in $\beta $ in the exponentials, we find that the new
transformed canonical variables $(\tilde{x}_{\mu },\tilde{p}_{\mu })$ obey
the brackets 
\begin{equation}
\{\tilde{p}_{\mu },\tilde{p}_{\nu }\}=(\beta -1)[\{p_{\mu },\beta \}p_{\nu
}+p_{\mu }\{\beta ,p_{\nu }\}]+\{\beta ,\beta \}p_{\mu }p_{\nu }  \tag{2.23a}
\end{equation}
\begin{equation*}
\{\tilde{x}_{\mu },\tilde{p}_{\nu }\}=(1+\beta )[\delta _{\mu \nu }(1-\beta
)-\{x_{\mu },\beta \}p_{\nu }]
\end{equation*}
\begin{equation}
+(1-\beta )x_{\mu }\{\beta ,p_{\nu }\}-\{\beta ,\beta \}x_{\mu }p_{\nu } 
\tag{2.23b}
\end{equation}
\begin{equation}
\{\tilde{x}_{\mu },\tilde{x}_{\nu }\}=(1+\beta )[x_{\mu }\{\beta ,x_{\nu
}\}-x_{\nu }\{\beta ,x_{\mu }\}]+\{\beta ,\beta \}x_{\mu }x_{\nu } 
\tag{2.23c}
\end{equation}
If we choose $\beta =\phi $ in equations (2.23) and compute the brackets on
the right side in terms of the Poisson brackets (2.10), we find the
expressions, after dropping terms proportional to $\beta ^{2}=\phi ^{2}$ 
\begin{equation}
\{\tilde{p}_{\mu },\tilde{p}_{\nu }\}=0  \tag{2.24a}
\end{equation}
\begin{equation}
\{\tilde{x}_{\mu },\tilde{p}_{\nu }\}=\eta _{\mu \nu }-p_{\mu }p_{\nu } 
\tag{2.24b}
\end{equation}
\begin{equation}
\{\tilde{x}_{\mu },\tilde{x}_{\nu }\}=-M_{\mu \nu }-M_{\mu \nu }\phi 
\tag{2.24c}
\end{equation}

Now, keeping the same order of approximation used to arrive at brackets
(2.23), that is, retaining only the linear terms in $\beta $, the
transformation equations (2.22a) and (2.22c) read 
\begin{equation}
\tilde{p}_{\mu }=\exp \{-\beta \}p_{\mu }=(1-\beta )p_{\mu }  \tag{2.25a}
\end{equation}
\begin{equation}
\tilde{x}_{\mu }=\exp \{\beta \}x_{\mu }=(1+\beta )x_{\mu }  \tag{2.25b}
\end{equation}
Using again the same function $\beta =\phi $ in equations (2.25), we write
them as 
\begin{equation}
\tilde{p}_{\mu }-p_{\mu }=c_{\mu }(x,p)\phi  \tag{2.26a}
\end{equation}
\begin{equation}
\tilde{x}_{\mu }-x_{\mu }=d_{\mu }(x,p)\phi  \tag{2.26b}
\end{equation}
where $c_{\mu }(x,p)=-p_{\mu }$ and $d_{\mu }(x,p)=x_{\mu }$. Equations
(2.26) are in the form (2.18) and so we can write 
\begin{equation}
\tilde{p}_{\mu }\approx p_{\mu }\text{ \ \ \ \ \ }\tilde{x}_{\mu }\approx
x_{\mu }  \tag{2.27}
\end{equation}
Using (2.18) and (2.27) in brackets (2.24), we can finally write the phase
space brackets 
\begin{equation}
\{p_{\mu },p_{\nu }\}\approx 0  \tag{2.28a}
\end{equation}
\begin{equation}
\{x_{\mu },p_{\nu }\}\approx \eta _{\mu \nu }-p_{\mu }p_{\nu }  \tag{2.28b}
\end{equation}
\begin{equation}
\{x_{\mu },x_{\nu }\}\approx -M_{\mu \nu }  \tag{2.28c}
\end{equation}
In a transition to the quantum theory by the correspondence principle rule
that [commutators]=$i\hbar $\{brackets\}, the brackets (2.28) will reproduce
the Snyder commutators [18] in the case when the noncommutativity parameter
is $\theta =1$. The Snyder commutators were obtained in a projective
geometry approach to the de Sitter space in the momentum representation.
Here we have derived their classical correspondents from the finite local
scale invariance (2.22) of the scalar massless particle Hamiltonian.
However, we have not succeeded in obtaining from the finite local scale
invariance (2.22) the Riemannian metric structure required by quantum
dynamics in the position representation. The massless particle does not have
enough gauge freedom for this metric structure to be derived in the same way
we derived the momentum space brackets (2.28). This is because the canonical
Hamiltonian (2.15) explicitly distinguishes momentum from position. As we
will see in the next section, this situation changes in 2T physics, where
momentum and position are indistinguishable variables, and a metric
structure can be derived in $d+2$ dimensions in exactly the same way we
derived the $d$ dimensional momentum space brackets (2.28).

It can be verified that brackets (2.28) satisfy all Jacobi identities among
the canonical variables, preserve the $d$ dimensional conformal algebra
(2.9) and preserve the first class property of constraint (2.17), therefore
preserving gauge invariance. Due to the non-trivial topology of the massless
particle configuration space, a vector field of vanishing strength tensor
must be present in the quantum theory. We will not consider this point here.
Instead we will concentrate on a discussion of this same situation in
classical 2T physics, which contains the $d$ dimensional massless scalar
relativistic particle as a gauge-fixed subsystem. For a general
parametrization of the classical solutions of 2T physics in any gauge, see
[6].

Hamiltonian (2.15) gives the classical equations of motion 
\begin{equation}
\dot{x}_{\mu }=\{x_{\mu },H\}=\lambda p_{\mu }  \tag{2.29a}
\end{equation}
\begin{equation}
\dot{p}_{\mu }=\{p_{\mu },H\}=0  \tag{2.29b}
\end{equation}
Equation (2.29b) shows that the massless particle moves with a constant
momentum relative to the parameter $\tau $, and is therefore a freely moving
particle. This situation changes in 2T physics because the $Sp(2,R)$ local
invariance or, in other words, the local indistinguishability between
position and momentum, brings with it an intrinsic interaction and as a
result a massless relativistic particle in a $d+2$ dimensional space-time
can no longer be completely free. The idea in this paper is that it feels
the effect of an intrinsic curved $d+2$ dimensional background.

\section{Two-time Physics}

In the usual one-time (1T) physics, a metric structure appears in the most
general configuration space formulation of quantum mechanics [1]. 1T physics
has been, and can always be used, to confirm the predictions of 2T physics.
In this paper we follow the opposite route. This route is to investigate the
possible existence of the $d+2$ dimensional generalization of a well known
situation in 1T physics. Specifically, here we are interested in the
construction of a $d+2$ dimensional general formulation of quantum
mechanics. This general formulation is expected to contain non relativistic
quantum mechanics in $d-1$ dimensions and relativistic quantum mechanics in $%
d$ dimensions as gauge-fixed subsectors. However, before trying to construct
such a theory, we must verify if its basic ingredients are available. In
this section we show how a natural metric structure can be found in $d+2$
dimensions. We start by reviewing the basic ideas that led to 2T physics.

The quantization rules of quantum mechanics are symmetric under the
interchange of coordinates and momenta. This is known as the discrete
symplectic symmetry $Sp(2)$ that transforms $(x,p)$ as a doublet. The
central idea in two-time physics [5-11] is to introduce a new gauge
invariance in phase space by gauging the duality of the quantum commutator $%
[X_{M},P_{N}]=i\hbar \eta _{MN}$. This procedure leads to a symplectic $%
Sp(2,R)$ gauge theory. To remove the distinction between position and
momentum we set $X_{1}^{M}=X^{M}$ and $X_{2}^{M}=P^{M}$ and define the
doublet $X_{i}^{M}=(X_{1}^{M},X_{2}^{M})$. The local $Sp(2,R)$ acts as 
\begin{equation}
\delta X_{i}^{M}(\tau )=\epsilon _{ik}\omega ^{kl}(\tau )X_{l}^{M}(\tau ) 
\tag{3.1}
\end{equation}
$\omega ^{ij}(\tau )$ is a symmetric matrix containing three local
parameters and $\epsilon _{ij}$ is the Levi-Civita symbol that serves to
raise or lower indices. The $Sp(2,R)$ gauge field $A^{ij}$ is symmetric in $%
(i,j)$ and transforms as 
\begin{equation}
\delta A^{ij}=\partial _{\tau }\omega ^{ij}+\omega ^{ik}\epsilon
_{kl}A^{lj}+\omega ^{jk}\epsilon _{kl}A^{il}  \tag{3.2}
\end{equation}
The covariant derivative is 
\begin{equation}
D_{\tau }X_{i}^{M}=\partial _{\tau }X_{i}^{M}-\epsilon _{ik}A^{kl}X_{l}^{M} 
\tag{3.3}
\end{equation}
An action invariant under the $Sp(2,R)$ gauge symmetry is 
\begin{equation}
S=\frac{1}{2}\int d\tau (D_{\tau }X_{i}^{M})\epsilon ^{ij}X_{j}^{N}\eta _{MN}
\tag{3.4}
\end{equation}
In Hamiltonian form action (3.4) becomes 
\begin{equation}
S=\int d\tau \lbrack \dot{X}.P-(\frac{1}{2}\lambda _{1}P^{2}+\lambda _{2}X.P+%
\frac{1}{2}\lambda _{3}X^{2})]  \tag{3.5}
\end{equation}
where $\lambda _{\alpha },$ $\alpha =1,2,3$ are Lagrange multipliers and the
canonical Hamiltonian is 
\begin{equation}
H=\frac{1}{2}\lambda _{1}P^{2}+\lambda _{2}X.P+\frac{1}{2}\lambda _{3}X^{2} 
\tag{3.6}
\end{equation}
The equations of motion for the $\lambda $'s give the first-class
constraints 
\begin{equation}
\phi _{1}=\frac{1}{2}P^{2}\approx 0  \tag{3.7}
\end{equation}
\begin{equation}
\phi _{2}=X.P\approx 0  \tag{3.8}
\end{equation}
\begin{equation}
\phi _{3}=\frac{1}{2}X^{2}\approx 0  \tag{3.9}
\end{equation}
Constraints (3.7)-(3.9), as well as evidences of two-time physics, were
independently obtained in [3]. The presence of first class constraints and
the associated gauge freedom indicates that there is more than one set of
canonical variables that corresponds to a given physical state [15].
However, equations (3.7) and (3.9) can be treated as constraints only if the
hypersurfaces $X_{0}=X_{1}=...=X_{d+1}=0$ and $P_{0}=P_{1}=...=P_{d+1}=0$
are excluded from phase space. Only in this case the gauge orbits generated
by $\phi _{1}$ and $\phi _{3}$ are regular [3,15]. Here then we also have a
phase space with a non-trivial topology. If we consider the Euclidean, or
the Minkowski metric as the background space-time, we find that the surface
defined by the constraint equations (3.7)-(3.9) is trivial. The only metric
giving a non-trivial surface, preserving the unitarity of the theory, and
avoiding the ghost problem is a flat metric with two time-like dimensions
[5-11]. Following [5-11] we introduce another space-like dimension and
another time-like dimension and start working in a Minkowski space with
signature $(d,2)$. Action (3.5) is the $d+2$ dimensional generalization of
the $d$ dimensional massless particle action (2.19). Action (3.5) describes
conformal gravity on the world-line [6,22,23]. Constraints (3.7)-(3.9) can
also be interpreted as describing a massless particle living on the border
of a $d+1$ dimensional AdS space of infinite radius [3].

In terms of the Poisson brackets 
\begin{equation}
\{P_{M},P_{N}\}=\{X_{M},X_{N}\}=0\text{ \ \ \ }\{X_{M},P_{N}\}=\eta _{MN} 
\tag{3.10}
\end{equation}
the local infinitesimal $Sp(2,R)$ transformations of action (3.5) are 
\begin{equation}
\delta X_{M}=\epsilon _{\alpha }(\tau )\{X_{M},\phi _{\alpha }\}=\epsilon
_{1}P_{M}+\epsilon _{2}X_{M}  \tag{3.11a}
\end{equation}
\begin{equation}
\delta P_{M}=\epsilon _{\alpha }(\tau )\{P_{M},\phi _{\alpha }\}=-\epsilon
_{2}P_{M}-\epsilon _{3}X_{M}  \tag{3.11b}
\end{equation}
\begin{equation}
\delta \lambda _{1}=\dot{\epsilon}_{1}+2\epsilon _{2}\lambda _{1}-2\epsilon
_{1}\lambda _{2}  \tag{3.11c}
\end{equation}
\begin{equation}
\delta \lambda _{2}=\dot{\epsilon}_{2}+\epsilon _{3}\lambda _{1}-\epsilon
_{1}\lambda _{3}  \tag{3.11d }
\end{equation}
\begin{equation}
\delta \lambda _{3}=\dot{\epsilon}_{3}+2\epsilon _{3}\lambda _{2}-2\epsilon
_{2}\lambda _{3}  \tag{3.11e}
\end{equation}
under which 
\begin{equation}
\delta S=\int_{\tau _{i}}^{\tau _{f}}d\tau \frac{d}{d\tau }(\epsilon
_{\alpha }\phi _{\alpha })  \tag{3.12}
\end{equation}
As in the massless particle case, since the interval $(\tau _{i},\tau _{f})$
is arbitrary, the quantity $Q=\epsilon _{\alpha }\phi _{\alpha }$ with $%
\alpha =1,2,3$ can be interpreted as the conserved Hamiltonian Noether
charge, or as the generator of the local infinitesimal transformations
(3.11), depending on wether the equations of motion are satisfied or not
[17].

Rigid infinitesimal $SO(d,2)$ transformations have the generator [5-11] 
\begin{equation}
L_{MN}=X_{M}P_{N}-X_{N}P_{M}  \tag{3.13}
\end{equation}
The $L_{MN}$ satisfy the algebra (2.12) and generate the transformations 
\begin{equation}
\delta X_{M}=-\frac{1}{2}\omega _{RS}\{X_{M},L_{RS}\}=\omega _{MR}X_{R} 
\tag{3.14a}
\end{equation}
\begin{equation}
\delta P_{M}=-\frac{1}{2}\omega _{RS}\{P_{M},L_{RS}\}=\omega _{MR}P_{R} 
\tag{3.14b}
\end{equation}
\begin{equation}
\delta \lambda _{\alpha }=0  \tag{3.14c}
\end{equation}
under which $\delta S=0$. Because the $L_{MN}$ are gauge invariant, $%
\{L_{MN},\phi _{\alpha }\}=0,$ the $SO(d,2)$ invariance is also present in
all the $d$ dimensional relativistic systems that can be obtained from the
2T physics action (3.5) by imposing two gauge conditions, and in all the $%
(d-1)$ dimensional non-relativistic systems that can be obtained from (3.5)
by imposing three gauge conditions.

Let us now consider how a Riemannian metric structure can be induced in the $%
d+2$ dimensional flat space-time of 2T physics. The 2T Hamiltonian (3.6) is
invariant under the finite local scale transformations 
\begin{equation}
\tilde{X}^{M}=\exp \{\beta (\tau )\}X^{M}  \tag{3.15a}
\end{equation}
\begin{equation}
\tilde{P}_{M}=\exp \{-\beta (\tau )\}P_{M}  \tag{3.15b}
\end{equation}
\begin{equation}
\tilde{\lambda}_{1}=\exp \{2\beta (\tau )\}\lambda _{1}  \tag{3.15c}
\end{equation}
\begin{equation}
\hat{\lambda}_{2}=\lambda _{2}  \tag{3.15d}
\end{equation}
\begin{equation}
\tilde{\lambda}_{3}=\exp \{-2\beta (\tau )\}\lambda _{3}  \tag{3.15e}
\end{equation}
where $\beta (\tau )$ is an arbitrary scalar function. The subsequent steps
are simply higher dimensional extensions of those for the massless particle.
Keeping only the linear terms in $\beta $ in transformation (3.15), we
arrive at the brackets 
\begin{equation}
\{\tilde{P}_{M},\tilde{P}_{N}\}=(\beta -1)[\{P_{M},\beta \}P_{N}+\{\beta
,P_{N}\}P_{M}]+\{\beta ,\beta \}P_{M}P_{N}  \tag{3.16a}
\end{equation}
\begin{equation*}
\{\tilde{X}_{M},\tilde{P}_{N}\}=(1+\beta )[\eta _{MN}(1-\beta
)-\{X_{M},\beta \}P_{N}]
\end{equation*}
\begin{equation}
+(1-\beta )X_{M}\{\beta ,P_{N}\}-X_{M}X_{N}\{\beta ,\beta \}  \tag{3.16b}
\end{equation}
\begin{equation}
\{\tilde{X}_{M},\tilde{X}_{N}\}=(1+\beta )[X_{M}\{\beta
,X_{N}\}-X_{N}\{\beta ,X_{M}\}]+X_{M}X_{N}\{\beta ,\beta \}  \tag{3.16c}
\end{equation}
If we choose $\beta =\phi _{1}$ in equations (3.16) and compute the brackets
on the right side using the Poisson brackets (3.10), we find the
expressions, after dropping terms proportional to $\beta ^{2}=\phi _{1}^{2}$ 
\begin{equation}
\{\tilde{P}_{M},\tilde{P}_{N}\}=0  \tag{3.17a}
\end{equation}
\begin{equation}
\{\tilde{X}_{M},\tilde{P}_{N}\}=\eta _{MN}-P_{M}P_{N}  \tag{3.17b}
\end{equation}
\begin{equation}
\{\tilde{X}_{M},\tilde{X}_{N}\}=-L_{MN}-L_{MN}\phi _{1}  \tag{3.17c}
\end{equation}

In the same order of approximation used to arrive at brackets (3.16),
transformation equations (3.15a) and (3.15b) read 
\begin{equation}
\tilde{X}^{M}=\exp \{\beta \}X^{M}=(1+\beta )X^{M}  \tag{3.18a}
\end{equation}
\begin{equation}
\tilde{P}_{M}=\exp \{-\beta \}P_{M}=(1-\beta )P_{M}  \tag{3.18b}
\end{equation}
Using again the same function $\beta =\phi _{1}$ in equations (3.18), we
write them as 
\begin{equation}
\tilde{X}^{M}-X^{M}=C_{\alpha }^{M}(X,P)\phi _{\alpha }  \tag{3.19a}
\end{equation}
\begin{equation}
\tilde{P}_{M}-P_{M}=D_{M}^{\alpha }(X,P)\phi _{\alpha }  \tag{3.19b}
\end{equation}
with $C_{1}^{M}=X^{M},$ $C_{2}^{M}=C_{3}^{M}=0$ and $D_{M}^{1}=-P_{M},$ $%
D_{M}^{2}=D_{M}^{3}=0$. Equations (3.19) are in the form (2.18) and so we
can write 
\begin{equation}
\tilde{X}^{M}\approx X^{M}\text{ \ \ \ \ \ \ \ }\tilde{P}_{M}\approx P_{M} 
\tag{3.20}
\end{equation}
Using these weak equalities in brackets (3.17) we can write the phase space
brackets

\begin{equation}
\{P_{M},P_{N}\}\approx 0  \tag{3.21a}
\end{equation}
\begin{equation}
\{X_{M},P_{N}\}\approx \eta _{MN}-P_{M}P_{N}  \tag{3.21b}
\end{equation}
\begin{equation}
\{X_{M},X_{N}\}\approx -L_{MN}  \tag{3.21c}
\end{equation}
Brackets (3.21) are the $(d+2)$ dimensional extensions of the $d$
dimensional momentum space brackets (2.28) we found for the massless
relativistic particle. But in 2T physics, where $X_{M}$ and $P_{M}$ are
locally indistinguishable variables, brackets (3.21) have a dual version in
position space. We can perform the duality transformation 
\begin{equation}
X_{M}\rightarrow P_{M}  \tag{3.22a}
\end{equation}
\begin{equation}
P_{M}\rightarrow -X_{M}  \tag{3.22b}
\end{equation}
\begin{equation}
\lambda _{1}\rightarrow \lambda _{3},\text{ \ }\lambda _{2}\rightarrow
-\lambda _{2},\text{ \ }\lambda _{3}\rightarrow \lambda _{1}  \tag{3.22c}
\end{equation}
which leaves the 2T Hamiltonian (3.6) invariant, and under which the 2T
action (3.5) transforms as $\delta S=-\int_{\tau _{i}}^{\tau _{f}}d\tau 
\frac{d}{d\tau }(X.P)$, being therefore invariant up to a surface term.
However, we can not simply substitute the duality transformations (3.22a)
and (3.22b) in bracket (3.21b) in order to obtain a metric structure in
position space in $d+2$ dimensions. This procedure introduces incorrect
minus signs in some of the resultant brackets and as a result some of the
Jacobi identities involving position and momentum fail to close. This is
because, as we saw in the introduction, the gravitational field, regarded as
a gauge field, corresponds to the group of continuous local scale
transformations and not to duality transformations of the type (3.22). The
correct procedure starts by noting that transformation (3.22) changes the
function $\beta =\frac{1}{2}P^{2},$ we used to arrive at brackets (3.21)
into the new function $\beta =\frac{1}{2}X^{2}.$ Introducing this new
function into brackets (3.16), which are the consequences in phase space of
the presence of the finite local scale invariance (3.15) of the 2T
Hamiltonian, and performing the same steps as in the case $\beta =\frac{1}{2}%
P^{2}$, we arrive at the position space brackets 
\begin{equation}
\{P_{M},P_{N}\}\approx L_{MN}  \tag{3.23a}
\end{equation}
\begin{equation}
\{X_{M},P_{N}\}\approx \eta _{MN}+X_{M}X_{N}  \tag{3.23b}
\end{equation}
\begin{equation}
\{X_{M},X_{N}\}\approx 0  \tag{3.23c}
\end{equation}
Notice that we can not obtain brackets (3.23) by performing the duality
transformation (3.22) in brackets (3.21). From equation (3.23b) we see that
we can use the finite local scale invariance (3.15) of the 2T Hamiltonian to
change from the flat Minkowski space with metric $\eta _{MN}$ to a
Riemannian space with metric tensor 
\begin{equation}
G_{MN}=\eta _{MN}+X_{M}X_{N}  \tag{3.24}
\end{equation}
This procedure of incorporating gravitational effects into quantum mechanics
by modifying the commutator $[X_{M},P_{N}]$ (or the corresponding classical
bracket, as is the case here) is not new and in the usual 1T physics it
becomes unavoidable [24] at energy scales near the Planck scale. In 2T
physics this procedure can not change the dynamic evolution of the system
because the Hamiltonian (3.6) is invariant under the local scale
transformation (3.15). In fact, Hamiltonian (3.6) generates the classical
equations of motion 
\begin{equation}
\dot{X}_{M}=\{X_{M},H\}=\lambda _{1}P_{M}+\lambda _{2}X_{M}  \tag{3.25a}
\end{equation}
\begin{equation}
\dot{P}_{M}=\{P_{M},H\}=-\lambda _{2}P_{M}-\lambda _{3}X_{M}  \tag{3.25b}
\end{equation}
computed in terms of the Poisson brackets (3.10). Equation (3.25b) shows
that the particle%
\'{}%
s momentum is no longer constant relative to the parameter $\tau $. An
interaction is perceived by the massless particle as a result of its
embedding in $d+2$ dimensions. The idea here is that it feels the effect of
the background (3.24).

It is easy to verify that if we leave the $d+2$ dimensional Minkowski space
of 2T physics, and use the local scale transformation (3.15) to change to
the $d+2$ dimensional space with metric tensor (3.24), the new Hamiltonian
will differ from (3.6) by terms that are quadratic in the first class
constraints. These quadratic terms can be dropped, and in the linear
approximation the Hamiltonian in the background (3.24) is identical to
(3.6). In addition, the equations of motion computed using the Hamiltonian
(3.6) and brackets (3.23) differ from the equations of motion (3.25) by
terms that are linear in the constraints. These linear terms can also be
dropped and the equations of motion in the background (3.24), computed in
terms of brackets (3.23), are identical to (3.25). We are then forced to
conclude that the $d+2$ dimensional space with metric tensor (3.24) is an
equally valid natural background for 2T physics because no homothety
condition [11] is necessary here. This was unknown until now. More
surprising is that, after dropping the terms proportional to the
constraints, the Hamiltonian and the equations of motion in the momentum
space background 
\begin{equation}
\bar{G}_{MN}=\eta _{MN}-P_{M}P_{N}  \tag{3.26}
\end{equation}
where brackets (3.21) are valid, are also identical to (3.6) and (3.25),
respectively. The flat $d+2$ dimensional Minkowski space is not the only
possible space for 2T physics. The $d+2$ dimensional position space (3.24)
and the $d+2$ dimensional momentum space (3.26) are also possible and more
general spaces for 2T physics. The important point here is that, reasoning
in analogy to known results in 1T physics [1], we may expect that the
transition to these more general $d+2$ dimensional backgrounds will be
necessary to guarantee the correct normalization of the position and
momentum eigenstates, the correct spectral decomposition of the identity
operator in the position and momentum eigenbasis, the correct matrix
elements of the position and momentum operators, and also the correct
integration measure for the inner product of any two states in a general
configuration space or momentum space formulation of quantum mechanics in $%
d+2$ dimensions.

To conclude this section we mention that in our derivation of brackets
(3.21) and (3.23) there is no need to use the Dirac bracket because there is
no second class constraint to begin with. Dirac brackets would have appeared
if we had imposed gauge conditions to turn the first class constraints
(3.7)-(3.9) into second class ones. This would bring us back to $d-1$
dimensions. An example of this is that the $d$ dimensional brackets (2.28),
we obtained for the massless particle using scale invariance arguments, can
also be derived as a Dirac bracket after imposing two canonical gauge
conditions [29] which turn the first class constraints (3.8) and (3.9) of 2T
physics into second class constraints. In this paper, this restriction of
the gauge freedom using the Dirac bracket technique, although possible, is
not necessary. This will guarantee that we are in $d+2$ dimensions.
Bypassing the Dirac brackets is a substantial advantage for our purposes
here. Indeed, the quantum realization of Dirac brackets that depend on the
canonical variables may be highly nontrivial\ and is by no means guaranteed
[15].

\section{2T Physics with Topological Vector Fields}

Now we explicitly take into account the non-trivial phase space topology of
2T physics. This can be done by introducing a vector field $A_{M}(X)$ which
defines a section of a flat U(1) bundle over space-time [1]. The vector
field must have a vanishing antisymmetric second rank strength tensor, $%
F_{MN}=\partial _{M}A_{N}-\partial _{N}A_{M}=0$. When this condition is met,
the flat U(1) bundle may be characterized [1], up to local infinitesimal
reparametrizations, by the differential 1-form $dX^{M}A_{M}(X).$

As we saw in section three, to obtain regular gauge orbits for the first
class constraints of 2T physics, the origin of phase space must be removed.
This creates a topological obstruction to the reduction of the vector field $%
A_{M}(X)$ to a pure gauge, $A_{M}=\frac{\partial \chi (X)}{\partial X^{M}}$,
where $\chi (X)$ is an arbitrary function. In other words, the vector field
must be present in the quantized 2T theory. In this section we search for a
corresponding classical action. As an initial attempt we modify action (3.5)
according to the usual minimal coupling prescription to vector fields, $%
P_{M}\rightarrow P_{M}-A_{M}$. This produces the correct U(1) covariant
derivative in the quantum theory [1]. The 2T action in this case is 
\begin{equation}
S=\int d\tau \{\dot{X}.P-[\frac{1}{2}\lambda _{1}(P-A)^{2}+\lambda
_{2}X.(P-A)+\frac{1}{2}\lambda _{3}X^{2}]\}  \tag{4.1}
\end{equation}
where the Hamiltonian is 
\begin{equation}
H=\frac{1}{2}\lambda _{1}(P-A)^{2}+\lambda _{2}X.(P-A)+\frac{1}{2}\lambda
_{3}X^{2}  \tag{4.2}
\end{equation}
The equations of motion for the multipliers now give the constraints 
\begin{equation}
\phi _{1}=\frac{1}{2}(P-A)^{2}\approx 0  \tag{4.3}
\end{equation}
\begin{equation}
\phi _{2}=X.(P-A)\approx 0  \tag{4.4}
\end{equation}
\begin{equation}
\phi _{3}=\frac{1}{2}X^{2}\approx 0  \tag{4.5}
\end{equation}
The Poisson brackets between the canonical variables and the vector field
are 
\begin{equation}
\{X_{M},A_{N}\}=0  \tag{4.6a}
\end{equation}
\begin{equation}
\{P_{M},A_{N}\}=-\frac{\partial A_{N}}{\partial X^{M}}  \tag{4.6b}
\end{equation}
\begin{equation}
\{A_{M},A_{N}\}=0  \tag{4.6c}
\end{equation}

Computing the algebra of constraints (4.3)-(4.5) using the Poisson brackets
(3.10) and (4.6) we obtain the equations 
\begin{equation}
\{\phi _{1},\phi _{1}\}=(P^{M}-A^{M})F_{MN}(P^{N}-A^{N})  \tag{4.7a}
\end{equation}
\begin{equation*}
\{\phi _{1},\phi _{2}\}=-2\phi _{1}+(P^{M}-A^{M})\frac{\partial }{\partial
X^{M}}(X.A)-(P-A).A
\end{equation*}
\begin{equation}
-X^{M}\frac{\partial }{\partial X^{M}}[(P-A).A]-X^{M}\frac{\partial }{%
\partial X^{M}}(\frac{1}{2}A^{2})  \tag{4.7b}
\end{equation}
\begin{equation}
\{\phi _{2},\phi _{2}\}=X^{M}F_{MN}X^{N}  \tag{4.7c}
\end{equation}
\begin{equation}
\{\phi _{1},\phi _{3}\}=-\phi _{2}  \tag{4.7d}
\end{equation}
\begin{equation}
\{\phi _{2},\phi _{3}\}=-2\phi _{3}  \tag{4.7e}
\end{equation}
\begin{equation}
\{\phi _{3},\phi _{3}\}=0  \tag{4.7f}
\end{equation}
For the case in which we are interested in this paper, we see from the above
equations that constraints (4.3)-(4.5) become first class constraints when
the vector field satisfies the conditions 
\begin{equation}
F_{MN}=0  \tag{4.8a}
\end{equation}
\begin{equation}
X.A=0  \tag{4.8b}
\end{equation}
\begin{equation}
(P-A).A=0  \tag{4.8c}
\end{equation}
\begin{equation}
\frac{1}{2}A^{2}=0  \tag{4.8d}
\end{equation}
Condition (4.8a) implies that the vector field $A_{M}(X)$ defines a section
of a flat U(1) bundle over the $d+2$ dimensional space-time. Observe that in
the case when $F_{MN}\neq 0$ the vanishing of bracket (4.7c) leads to the
same condition (1.4) obtained in [11]. But here a careful look at bracket
(4.7a) suggests that, in the presence of a vector field for which $%
F_{MN}\neq 0$, condition (1.4) should be complemented with the condition $%
(P^{M}-A^{M})F_{MN}=0$. This would render the theory simultaneously in
agreement with the minimal coupling prescription to vector fields and with
the local indistinguishability between $X^{M}$ and $P^{M}-A^{M}(X)$ in the
presence of the vector field. A curious observation is that $P^{M}$ also
becomes indistinguishable from $X^{M}-A^{M}(P).$ This point will be
considered in a future paper [21].

As can be easily verified, conditions (4.8b)-(4.8d) imply that constraints
(4.3)-(4.5) are not the irreducible [15] set of constraints for 2T physics
with a topological vector field. Combining then conditions (4.8b)-(4.8d)
with constraints (4.3)-(4.5), we obtain the irreducible set of constraints 
\begin{equation}
\phi _{1}=\frac{1}{2}P^{2}\approx 0  \tag{4.9}
\end{equation}
\begin{equation}
\phi _{2}=X.P\approx 0  \tag{4.10}
\end{equation}
\begin{equation}
\phi _{3}=\frac{1}{2}X^{2}\approx 0  \tag{4.11}
\end{equation}
\begin{equation}
\phi _{4}=X.A\approx 0  \tag{4.12}
\end{equation}
\begin{equation}
\phi _{5}=P.A\approx 0  \tag{4.13}
\end{equation}
\begin{equation}
\phi _{6}=\frac{1}{2}A^{2}\approx 0  \tag{4.14}
\end{equation}
Observe that Dirac's conditions (1.2a) and (1.2b) are now reproduced by
constraints $\phi _{4}$ and $\phi _{5}$. The contrast with the set (1.2) is
that our calculation leads to a scalar third condition on the vector field,
a condition which will now be verified to be the correct constraint for the
2T theory in the presence of a vector field for which $F_{MN}=0$.

It can be verified that constraints (4.9)-(4.14) are all first class. We can
then write down the Hamiltonian action 
\begin{equation*}
S=\int_{\tau _{i}}^{\tau _{f}}d\tau \lbrack \dot{X}.P-(\frac{1}{2}\lambda
_{1}P^{2}+\lambda _{2}X.P+\frac{1}{2}\lambda _{3}X^{2}
\end{equation*}
\begin{equation}
+\lambda _{4}X.A+\lambda _{5}P.A+\frac{1}{2}\lambda _{6}A^{2})]  \tag{4.15}
\end{equation}
describing two-time physics with a vector field of topological origin. The
Hamiltonian is 
\begin{equation*}
H=\frac{1}{2}\lambda _{1}P^{2}+\lambda _{2}X.P+\frac{1}{2}\lambda _{3}X^{2}.
\end{equation*}
\begin{equation}
+\lambda _{4}X.A+\lambda _{5}P.A+\frac{1}{2}\lambda _{6}A^{2}  \tag{4.16}
\end{equation}

The $L_{MN}$ in (3.13) generate the rigid infinitesimal $SO(d,2)$
transformations in action (4.15) 
\begin{equation}
\delta X_{M}=-\frac{1}{2}\omega _{RS}\{X_{M},L_{RS}\}=\omega _{MR}X_{R} 
\tag{4.17a}
\end{equation}
\begin{equation}
\delta P_{M}=-\frac{1}{2}\omega _{RS}\{P_{M},L_{RS}\}=\omega _{MR}P_{R} 
\tag{4.17b}
\end{equation}
\begin{equation}
\delta A_{M}=\frac{\partial A_{M}}{\partial X_{R}}\delta X_{R}.  \tag{4.17c}
\end{equation}
\begin{equation}
\delta \lambda _{\varrho }=0,\text{ \ }\varrho =1,2,...,6  \tag{4.17d}
\end{equation}
under which $\delta S=0.$ It can be checked that $L_{MN}$ has weakly
vanishing brackets with the first class constraints (4.9)-(4.14), being
therefore gauge invariant.

Action (4.15) also has the local infinitesimal invariance 
\begin{equation}
\delta X_{M}=\epsilon _{\varrho }(\tau )\{X_{M},\phi _{\varrho }\}=\epsilon
_{1}P_{M}+\epsilon _{2}X_{M}+\epsilon _{5}A_{M}  \tag{4.18a}
\end{equation}
\begin{equation*}
\delta P_{M}=\epsilon _{\varrho }(\tau )\{P_{M},\phi _{\varrho }\}=-\epsilon
_{2}P_{M}-\epsilon _{3}X_{M}-\epsilon _{4}A_{M}
\end{equation*}
\begin{equation}
-\epsilon _{4}X_{N}\frac{\partial A_{N}}{\partial X^{M}}-\epsilon _{5}P_{N}%
\frac{\partial A_{N}}{\partial X^{M}}-\epsilon _{6}A_{N}\frac{\partial A_{N}%
}{\partial X^{M}}  \tag{4.18b}
\end{equation}
\begin{equation}
\delta A_{M}=\frac{\partial A_{M}}{\partial X^{N}}\delta X^{N}  \tag{4.18c}
\end{equation}
\begin{equation}
\delta \lambda _{1}=\dot{\epsilon}_{1}+2\epsilon _{2}\lambda _{1}-2\epsilon
_{1}\lambda _{2}  \tag{4.18d}
\end{equation}
\begin{equation}
\delta \lambda _{2}=\dot{\epsilon}_{2}+\epsilon _{3}\lambda _{1}-\epsilon
_{1}\lambda _{3}  \tag{4.18e}
\end{equation}
\begin{equation}
\delta \lambda _{3}=\dot{\epsilon}_{3}+2\epsilon _{3}\lambda _{2}-2\epsilon
_{2}\lambda _{3}  \tag{4.18f}
\end{equation}
\begin{equation}
\delta \lambda _{4}=\dot{\epsilon}_{4}+\epsilon _{3}\lambda _{5}-\epsilon
_{5}\lambda _{3}  \tag{4.18g}
\end{equation}
\begin{equation}
\delta \lambda _{5}=\dot{\epsilon}_{5}+\epsilon _{2}\lambda _{5}-\epsilon
_{5}\lambda _{2}  \tag{4.18h}
\end{equation}
\begin{equation}
\delta \lambda _{6}=\dot{\epsilon}_{6}  \tag{4.18i}
\end{equation}
under which 
\begin{equation}
\delta S=\int_{\tau _{i}}^{\tau _{f}}d\tau \frac{d}{d\tau }(\epsilon _{\rho
}\phi _{\rho })  \tag{4.19}
\end{equation}
Now the conserved charge, or the generator of the local transformations,
depending on wether the equations of motion are satisfied or not, is the
quantity $Q=\epsilon _{\rho }\phi _{\rho }$ with $\rho =1,2,...,6$ \ This
generalizes the local infinitesimal invariance (3.11) of 2T physics to the
case when a vector field of vanishing strength tensor is present.

Hamiltonian (4.16) is invariant under the finite local scale transformations 
\begin{equation}
\tilde{X}^{M}=\exp \{\beta (\tau )\}X^{M}  \tag{4.20a}
\end{equation}
\begin{equation}
\tilde{P}_{M}=\exp \{-\beta (\tau )\}P_{M}  \tag{4.20b}
\end{equation}
\begin{equation}
\tilde{A}_{M}=\exp \{-\beta (\tau )\}A_{M}  \tag{4.20c}
\end{equation}
\begin{equation}
\tilde{\lambda}_{1}=\exp \{2\beta (\tau )\}\lambda _{1}  \tag{4.20d}
\end{equation}
\begin{equation}
\tilde{\lambda}_{2}=\lambda _{2}  \tag{4.20e}
\end{equation}
\begin{equation}
\tilde{\lambda}_{3}=\exp \{-2\beta (\tau )\}\lambda _{3}  \tag{4.20f}
\end{equation}
\begin{equation}
\tilde{\lambda}_{4}=\lambda _{4}  \tag{4.20g}
\end{equation}
\begin{equation}
\tilde{\lambda}_{5}=\exp \{2\beta (\tau )\}\lambda _{5}  \tag{4.20h}
\end{equation}
\begin{equation}
\tilde{\lambda}_{6}=\exp \{2\beta (\tau )\}\lambda _{6}  \tag{4.20i}
\end{equation}
Notice that transformations (4.20b) and (4.20c) are consistent with the
minimal coupling prescription to vector fields. Using the invariance (4.20)
we can arrive, if we choose $\beta =\phi _{1},$ at the same brackets (3.21).
We can arrive at brackets (3.23) by choosing $\beta =\phi _{3}$ as before.
The metric structure (3.24) in position space and the metric structure
(3.26) in momentum space we obtained in section three are then both
preserved in the presence of a vector field of vanishing strength tensor.

In the presence of the vector field we can again change to the backgrounds
(3.24) or (3.26) without changing the dynamic evolution of the system. For
instance, if we perform the change to the background (3.24) we find that the
new Hamiltonian differs from (4.16) by terms that are quadratic in the first
class constraints (4.9)-(4.14). These quadratic terms can again be dropped
and in the linear approximation the Hamiltonian in the background (3.24),
and in the presence of the vector field, is identical to (4.16). The
classical equations of motion\ generated by the Hamiltonian\ (4.16),
computed in terms of the Poisson brackets (3.10) and (4.6), are 
\begin{equation}
\dot{X}_{M}=\{X_{M},H\}=\lambda _{1}P_{M}+\lambda _{2}X_{M}+\lambda _{5}A_{M}
\tag{4.21a}
\end{equation}
\begin{equation*}
\dot{P}_{M}=\{P_{M},H\}=-\lambda _{2}P_{M}-\lambda _{3}X_{M}-\lambda
_{4}A_{M}
\end{equation*}
\begin{equation}
-\lambda _{4}X_{N}\frac{\partial A_{N}}{\partial X^{M}}-\lambda _{5}P_{N}%
\frac{\partial A_{N}}{\partial X^{M}}-\lambda _{6}A_{N}\frac{\partial A_{N}}{%
\partial X^{M}}  \tag{4.21b}
\end{equation}
\begin{equation}
\dot{A}_{M}=\{A_{M},H\}=\lambda _{1}P_{N}\frac{\partial A_{M}}{\partial X^{N}%
}+\lambda _{2}X_{N}\frac{\partial A_{M}}{\partial X^{N}}+\lambda _{5}A_{N}%
\frac{\partial A_{M}}{\partial X^{N}}  \tag{4.21c}
\end{equation}
The local scale transformation (4.20) with the function $\beta =\frac{1}{2}%
X^{2}$ changes the Poisson brackets (4.6) into the new set 
\begin{equation}
\{X_{M},A_{N}\}=0  \tag{4.22a}
\end{equation}
\begin{equation}
\{P_{M},A_{N}\}=-\frac{\partial A_{N}}{\partial X^{M}}+X_{M}A_{N} 
\tag{4.22b}
\end{equation}
\begin{equation}
\{A_{M},A_{N}\}=0  \tag{4.22c}
\end{equation}
Now, computing the equations of motion generated by the Hamiltonian (4.16)
in terms of the brackets (3.23) and (4.22), we find that these equations
differ from equations (4.21) by terms that are linear in the first class
constraints (4.9)-(4.14). These terms can be dropped and the classical
equations of motion in the background (3.24) become identical to (4.21),
which are valid in the flat $d+2$ dimensional background. The same situation
occurs in the background (3.26) after the Poisson brackets (4.6) are
replaced by the brackets 
\begin{equation}
\{X_{M},A_{N}\}=-P_{M}A_{N}-X_{M}P_{S}\frac{\partial A_{N}}{\partial X^{S}} 
\tag{4.23a}
\end{equation}
\begin{equation}
\{P_{M},A_{N}\}=-\frac{\partial A_{N}}{\partial X^{M}}+P_{M}P_{S}\frac{%
\partial A_{N}}{\partial X^{S}}  \tag{4.23b}
\end{equation}
\begin{equation}
\{A_{M},A_{N}\}=A_{M}P_{S}\frac{\partial A_{N}}{\partial X^{S}}-A_{N}P_{S}%
\frac{\partial A_{M}}{\partial X^{S}}  \tag{4.23c}
\end{equation}
which emerge after the local scale transformation (4.20) with $\beta =\frac{1%
}{2}P^{2}$ is performed. Also in the presence of the vector field, the $d+2$
dimensional Minkowski space is not the only possible space for 2T physics.
The $d+2$ dimensional spaces given by (3.24) and (3.26) are also possible
spaces. Although these three spaces are indistinguishable at the classical
level, this situation may change in quantum mechanics because the correct
quantum dynamics may emerge only after these underlying metric structures in
position and momentum spaces, together with the vector field of vanishing
strength tensor, are explicitly taken into account in all the relevant
equations.

\section{Concluding remarks}

In this paper we showed that it is possible to construct, in the $d+2$
dimensional space-time of classical 2T physics, the same geometrical and
topological structures that are present in the most general configuration
space formulation of quantum mechanics containing gravity in $d$ dimensions.
The geometric structure is defined by a symmetric Riemannian metric tensor
and the topological structure is defined by a vector field with a vanishing
antisymmetric strength tensor which defines a section of a flat U(1) bundle
over space-time. This $d+2$ dimensional construction is possible first
because of the existence of a finite local scale invariance of the 2T
canonical Hamiltonian, and second because 2T physics contains at the
classical level a local continuous generalization of the discrete duality
symmetry between position and momentum that underlies the structure of
quantum mechanics.

One of the results of this paper that\ requires a deeper investigation is
the fact that the classical Hamiltonian 2T dynamics in the presence of the
topological vector field and described by the variables $X_{M}$, $P_{M}$ and 
$A_{M}(X)$ satisfying the brackets 
\begin{equation*}
\{P_{M},P_{N}\}=0
\end{equation*}
\begin{equation*}
\{X_{M},P_{N}\}=\eta _{MN}
\end{equation*}
\begin{equation*}
\{X_{M},X_{N}\}=0
\end{equation*}
\begin{equation*}
\{X_{M},A_{N}\}=0
\end{equation*}
\begin{equation*}
\{P_{M},A_{N}\}=-\frac{\partial A_{N}}{\partial X^{M}}
\end{equation*}
\begin{equation*}
\{A_{M},A_{N}\}=0
\end{equation*}
is the same classical Hamiltonian dynamics described by the same variables
but satisfying the brackets 
\begin{equation*}
\{P_{M},P_{N}\}=L_{MN}
\end{equation*}
\begin{equation*}
\{X_{M},P_{N}\}=\eta _{MN}+X_{M}X_{N}
\end{equation*}
\begin{equation*}
\{X_{M},X_{N}\}=0
\end{equation*}
\begin{equation*}
\{X_{M},A_{N}\}=0
\end{equation*}
\begin{equation*}
\{P_{M},A_{N}\}=-\frac{\partial A_{N}}{\partial X^{M}}+X_{M}A_{N}
\end{equation*}
\begin{equation*}
\{A_{M},A_{N}\}=0
\end{equation*}
and is also the same classical Hamiltonian dynamics described by the same
variables but now satisfying the brackets 
\begin{equation*}
\{P_{M},P_{N}\}=0
\end{equation*}
\begin{equation*}
\{X_{M},P_{N}\}=\eta _{MN}-P_{M}P_{N}
\end{equation*}
\begin{equation*}
\{X_{M},X_{N}\}=-L_{MN}
\end{equation*}
\begin{equation*}
\{X_{M},A_{N}\}=-P_{M}A_{N}-X_{M}P^{S}\frac{\partial A_{N}}{\partial X^{S}}
\end{equation*}
\begin{equation*}
\{P_{M},A_{N}\}=-\frac{\partial A_{N}}{\partial X^{M}}+P_{M}P^{S}\frac{%
\partial A_{N}}{\partial X^{S}}
\end{equation*}
\begin{equation*}
\{A_{M},A_{N}\}=A_{M}P^{S}\frac{\partial A_{N}}{\partial X^{S}}-A_{N}P^{S}%
\frac{\partial A_{M}}{\partial X^{S}}
\end{equation*}
We may say that, as a consequence of finite local scale invariance, three
formulations of quantum dynamics in three different spaces have the same
classical Hamiltonian limit described by 2T physics.

Inspired by the example of the spherical harmonic oscillator in a punctured
plane described in [1], we are inclined to look at the holonomy parametrized
by the 1-form $dX^{M}A_{M}$ as a higher dimensional Aharonov-Bohm flux line
[25] piercing the configuration space at its origin and in whose vector
potential the $d+2$ dimensional massless particle moves. This can explain
the noncommutativity of the momenta that will be induced in the quantized
theory if bracket (3.23a) is assumed. It is well known [26] that in a
magnetic field the momenta fail to mutually commute. However, the vector
field considered in this paper should not necessarily be interpreted as
having an electrodynamic origin because no electric charge was assumed for
the particle. Also, as became clear in the ADM construction [27] of general
relativity, a neutral scalar massless relativistic particle couples only to
the gravitational field. A gravitational or topological interpretation for
the vector field is then also possible. An exotic but interesting
possibility may be to interpret the vector field as having a gravitodynamic
origin [28]. In any case, it turns out that in $d+2$ dimensions the
nontrivial holonomies associated to the nontrivial representations of the
Heisenberg algebra can also be regarded as being due to some specific
Aharonov-Bohm flux lines passing through holes in configuration space, and
which are characterized by the first homotopy group $\pi _{1}(M)$ of that
space.

To conclude we would like to mention that the main motivation for this paper
is to try to apply the ideas of 2T physics in gravitational and quantum
mechanical physics. This area of theoretical physics seems to be left rather
unexplored by the main researchers in 2T physics. However, it is now clear
that the Standard Model of Particles and Forces in 3+1 dimensions is only
part of a "master 2T theory" [30] in 4+2 dimensions. This "master 2T theory"
is exactly the massless particle in flat $d+2$ dimensions described by
action (3.5) in the case when $d=4$. The results of this paper then have the
potential to bring with them entirely new ways of incorporating
gravitational and topological effects into the Standard Model. As emphasized
in [30], the higher space in $d+2$ dimensions is not just formalism that
could be avoided. 1T physics can be used to verify and interpret the
predictions of 2T physics, but it is not equipped to come up with the
predictions in the first place [30], unless one stumbles into some of them
occasionally, such as the $SO(d,2)$ conformal symmetry of the massless
scalar relativistic particle we considered in section two. The results in
this paper are then relevant because they teach us that one of the great
advantages of 2T physics over 1T physics is the classical continuous local
indistinguishability of position and momenta it explicitly displays. Before
the advent of 2T physics this kind of indistinguishability, but in a much
more restricted discontinuous global form, was long known to exist in
quantum mechanics as a consequence of the wave-particle duality of matter
and energy. The lessons from 2T physics so far makes it evident that the
ordinary 1T physics formulation of Nature is insufficient to provide the
explanation or even the existence [30] of the many unifying facts revealed
through 2T physics.

\noindent

\noindent


\bigskip

\begin{thebibliography}{99}
\bibitem{1}  J. Govaerts and V. Villanueva, Inter. Jour. Mod. Phys. A15
(2000) 4903 (arXiv:quant-ph/9908014)

\bibitem{2}  E. Witten, Adv. Theor. Math. Phys. 2 (1998) 253
(arXiv:hep-th/9802150)

\bibitem{3}  C. Leiva and M. S. Plyushchay, Ann. Phys. 307 (2003) 372
(arXiv:hep-th/0301244)

\bibitem{4}  J. M. Maldacena, Adv. Theor. Math. Phys. 2 (1998) 231; J. M.
Maldacena, Int. J. Theor. Phys.38 (1999) 1113 (arXiv:hep-th/9711200)

\bibitem{5}  \noindent I. Bars and C. Kounnas, Phys. Lett. B402 (1997) 25
(arXiv:hep-th/9703060)

\bibitem{6}  I. Bars, C. Deliduman and O. Andreev, Phys. Rev. D58 (1998)
066004 (arXiv:hep-th/9803188)

\bibitem{7}  I. Bars and C. Kounnas, Phys. Rev. D56 (1997) 3664
(arXiv:hep-th/9705205)

\bibitem{8}  I. Bars, arXiv:hep-th/9809034

\bibitem{9}  I. Bars, C. Deliduman and D. Minic, Phys. Lett. B 466 (1999)
135 (arXiv:hep-th/9906223)

\bibitem{10}  I. Bars, Phys. Rev. D58 (1998) 066006 (arXiv:hep-th/9804028)

\bibitem{11}  I. Bars, Phys. Rev. D62 (2000) 085015 (arXiv:hep-th/0002140)

\bibitem{12}  P. A. M. Dirac, Ann. Math. 37 (1936) 429

\bibitem{13}  V. I. Arnold, \textsl{Mathematical Methods of Classical
Mechanics}, Mir Editions 1979, Moscow

\bibitem{14}  P. A. M. Dirac, \textsl{Lectures on Quantum Mechanics, }%
Yeshiva University, 1964

\bibitem{15}  M. Henneaux and C. Teitelboim, \textsl{Quantization of Gauge
Systems,} Princeton University Press 1992

\bibitem{16}  M. S. Plyushchay, Nucl. Phys. B589 (2000) 413
(arXiv:hep-th/0004032)

\bibitem{17}  V. M. Villanueva, J. A. Nieto, L. Ruiz and J. Silvas, Jour.
Phys. A38 (2005) 7183 (arXiv:hep-th/0503093)

\bibitem{18}  H. S. Snyder, Phys. Rev. 71 (1947) 38

\bibitem{19}  W. Chagas-Filho, \textsl{Dual Gravitation},\
arXiv:hep-th/0611118

\bibitem{20}  N. P. Konopleva and V. N. Popov, \textsl{Gauge Fields},
Harwood Academic Publishers, 1981

\bibitem{21}  W. Chagas-Filho, \textsl{2T Physics and Quantum Mechanics, }to
appear in arXiv:hep-th

\bibitem{22}  R. Marnelius, Phys. Rev. D20 (1979) 2091

\bibitem{23}  W. Siegel, Int. J. Mod. Phys. A3 (1988) 2713

\bibitem{24}  D. V. Ahluwalia, Phys. Lett. B339 (1994) 301
(arXiv:hep-th/9308007)

\bibitem{25}  Y. Aharonov and D. Bohm, Phys. Rev. 115 (1959) 485

\bibitem{26}  M. Douglas and N. Nekrasov, Rev. Mod. Phys. 73 (2001) 977
(arXiv:hep-th/0106048)

\bibitem{27}  R. Arnowitt, S. Deser and C. W. Misner, Phys. Rev. 117 (1960)
1595 (arXiv:gr-qc/0405109)

\bibitem{28}  H. Behera, arXiv:gr-qc/0510003

\bibitem{29}  J. M. Romero and A. Zamora, Phys. Rev. D70 (2004) 105006,
(arXiv:hep-th/0408193)

\bibitem{30}  I. Bars, S-H. Chen and G. Qu\'{e}lin, arXiv:hep-th0705.2834
\end{thebibliography}
\end{document}